# Boosting Moving Object Indexing through Velocity Partitioning

Thi Nguyen [#1], Zhen He [#2], Rui Zhang [*3], Phillip Ward [#†4]

[#]*Department of Computer Science and Computer Engineering, La Trobe University, Australia*
[1]`nt2nguyen@students.latrobe.edu.au`, [2]`z.he@latrobe.edu.au`

[*]*Department of Computing and Information Systems, University of Melbourne, Australia*
[3]`rui@csse.unimelb.edu.au`

[†]*CSIRO Land and Water, Highett, Victoria, Australia*
[4]`p.ward@csiro.au`

## ABSTRACT

There have been intense research interests in moving object indexing in the past decade. However, existing work did not exploit the important property of skewed velocity distributions. In many real world scenarios, objects travel predominantly along only a few directions. Examples include vehicles on road networks, flights, people walking on the streets, etc. The search space for a query is heavily dependent on the velocity distribution of the objects grouped in the nodes of an index tree. Motivated by this observation, we propose the *velocity partitioning (VP)* technique, which exploits the skew in velocity distribution to speed up query processing using moving object indexes. The VP technique first identifies the "dominant velocity axes (DVAs)" using a combination of principal components analysis (PCA) and $k$-means clustering. Then, a moving object index (e.g., a TPR-tree) is created based on each DVA, using the DVA as an axis of the underlying coordinate system. An object is maintained in the index whose DVA is closest to the object's current moving direction. Thus, all the objects in an index are moving in a near 1-dimensional space instead of a 2-dimensional space. As a result, the expansion of the search space with time is greatly reduced, from a quadratic function of the maximum speed (of the objects in the search range) to a near linear function of the maximum speed. The VP technique can be applied to a wide range of moving object index structures. We have implemented the VP technique on two representative ones, the TPR*-tree and the $B^x$-tree. Extensive experiments validate that the VP technique consistently improves the performance of those index structures.

## 1. INTRODUCTION

GPS enabled mobile devices (phones, car navigators, etc) are ubiquitous these days and it is common for them to report their locations to a server in order to get location based services. Such services involve querying the current or near future locations of the mobile devices. Many index structures have been proposed to facilitate efficient query processing on moving objects in the last decade (e.g., [8, 13, 17, 20, 21, 23, 25]). However, none of these index structures exploit the important property of skewed velocity distributions. In most real world scenarios, objects travel predominantly along only a few directions due to the fixed underlying trav-elling infrastructure or routes. Examples include vehicles on road networks, flights, people walking on the streets, etc. Figure 1(a) shows a portion of the road network of San Francisco, where most of the roads are along two directions. Figure 1(b) shows a sample of velocity distribution of the cars travelling on the San Francisco road network. Every point (2-dimensional vector) in the figure represents the velocity of a car. It is clear that most of the cars are travelling along two dominant directions (axes).

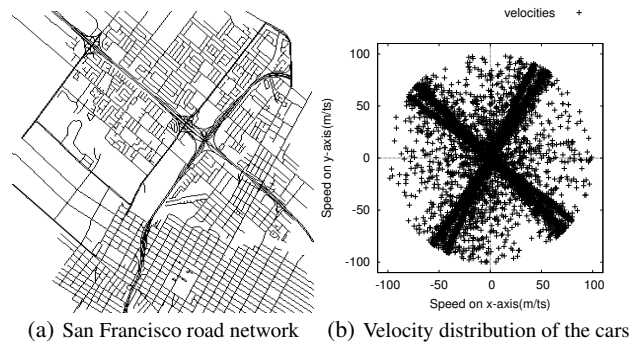

(a) San Francisco road network   (b) Velocity distribution of the cars

**Figure 1: San Francisco road network and the cars' velocity distribution**

The velocity distribution of objects in an index has a great impact on the rate at which the query search space expands. The search space expansion is either due to the tree nodes' minimum bounding rectangle (MBR) expansion (e.g., the TPR-tree/TPR*-tree [21, 23]) or query expansion (e.g., the $B^x$-tree [13]). In either case, the search space for a tree node is enlarged during the query time interval using the largest speed of the objects grouped in that tree node. If the velocities of the objects in a node are randomly distributed, then the search space is enlarged along both the $x$- and $y$-axes, and therefore there is a quadratic function of the maximum speed of the objects in the node. If the movements of all the objects in a node are largely along the same direction, then the search space is enlarged mainly along one axis and hence there is close to a linear function of the maximum speed of the objects in the node.

Motivated by this observation, we propose the *velocity partitioning (VP)* technique, which exploits the skew in velocity distribution to speed up query processing using moving object indexes. The VP technique first identifies the "dominant velocity axes (DVAs)" using a combination of principal components analysis (PCA) and $k$-means clustering. A DVA is an axis, which the velocities of most of the objects are (almost) parallel to. Then, a moving object index (e.g., a TPR-tree) is created based on each DVA, using the DVA as an axis of the underlying coordinate system. Objects are dynamically moved between DVA indexes when their movement directions change from one DVA to another. Objects with current velocities, which are far from any DVAs, are put in an outlier index. The outlier index uses the regular coordinate system. Thus, except for the outlier index, the objects in each other index are moving in a near 1-dimensional space instead of a 2-dimensional space. As a result, the expansion of the search space with time is greatly reduced, from a quadratic function of the maximum speed (of the objects in the search range) to a near linear function of the maximum speed.

The VP technique is a generic method and can be applied to a wide range of moving object index structures. In this paper, we focus our analysis and implementation of the VP technique on the two most well recognized and representative moving object indexes of different styles, the TPR*-tree [23] and the $B^x$-tree [13]. These two indexes are the basis for many recent indexing techniques [7, 22, 24, 25]. Our method can be applied to these more recent indexes in similar ways to how it is applied to those two representative indexes. We perform an extensive set of experiments using various real and synthetic data sets. The results show that the VP technique consistently improves the performance of both index structures. The improvement is up to around 3 times in terms of both query I/O and query execution time for both index structures.

The contributions of this paper are summarized below:

- We analytically show why a moving object index with VP outperforms a moving object index without VP.
- We propose the VP technique, which identifies the dominant velocity axes (DVAs) and maintain the objects in separate indexes based on the DVAs.
- We analytically show how to choose the value of an important parameter that determines which objects belong to the outlier index.
- We implemented the VP technique on two state-of-the-art moving object indexes, the TPR*-tree and the $B^x$-tree. We have performed an extensive experimental study. The results validate the effectiveness of our approach across a large number of real and synthetic data sets.

## 2. PRELIMINARIES

In this section, we provide some background on moving objects, and briefly review two techniques used in our approach, principal components analysis (PCA) and $k$-means clustering.

### 2.1 Moving Object Representation and Querying

A simple way of tracking the location of moving objects is to take location samples periodically. However, this approach requires frequent location updates, which imposes a heavy workload on the system. A popular method to reduce the reporting rate is to use a linear function to describe the near future trajectory of moving objects. The model consists of the initial location of the object and a velocity vector. An update is issued by the object when its velocity changes. An object velocity update simply consists of a deletion followed by an insertion. This linear model based approach is used by many studies [8, 13, 17, 19, 20, 21, 23, 25, 26, 28] on indexing and querying moving objects. We also follow this model in this paper, and the moving objects are modeled as moving points.

We support three different types of range queries: *time slice range query*, which reports the objects within the query range at a particular time stamp; *time interval range query*, which reports the objects within the query range within a time range; *moving range query*, where the query range itself is moving and the query reports the objects that intersect the moving range in a time range. For all three types of range queries, if the query timestamp (or time range) is in the future, the query range is projected (expanded) to that future time to check which objects should be returned.

### 2.2 Principal Components Analysis

Principal components analysis (PCA) is a commonly used method for *dimensionality reduction* [4, 12] and for finding correlations among attributes of data [15]. It examines the variance structure in the data set and determines the directions along which the data exhibits high variance. In our case, if we map the velocity of objects into the 2D velocity space as points, then the axis with high variance is the DVA.

Given a set of $k$-dimensional data points, PCA finds a ranked set of orthogonal $k$-dimensional eigenvectors $v_1, v_2, ..., v_k$ (which we call principal component vectors) such that:

- Each principal component (PC) vector is a unit vector, i.e., $\sqrt{\beta_{i_1}^2 + \beta_{i_2}^2 + ... + \beta_{i_k}^2} = 1$, where $\beta_{ij}$ ($i, j = 1,2, ...,k$) is the $j^{th}$ component of the PC vector $v_i$.
- The first PC $v_1$ accounts for most of the variability in the data, and each succeeding component accounts for as much of the remaining variability as possible.

### 2.3 $K$-means Clustering

$K$-means clustering [18] is a method commonly used to automatically partition a data set into $k$ clusters where each data point belongs to the cluster with the nearest *centroid*. It starts by assigning each object to one of $k$ clusters either randomly or using some heuristic method. The centroid of each cluster is computed and each point is re-assigned to its closest cluster centroid. When all points have been assigned, the $k$ cluster centroids are recomputed. The process is repeated until the centroids no longer move.

## 3. RELATED WORK

In this section, we review existing work on moving object indexes, specifically R-tree [3] based indexes, the $B^x$-tree [13], and dual transform based indexes. We also discuss indexing techniques for handling skewed workloads and for handling moving objects on road networks.

### 3.1 R-tree Based Moving Object Indexes

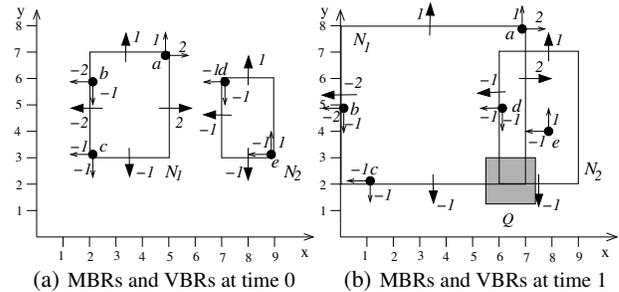

(a) MBRs and VBRs at time 0    (b) MBRs and VBRs at time 1

**Figure 2: MBRs of a TPR-tree growing with time**

An established approach to index moving objects is to use the R-tree [3] or it's more optimized variant the R*-tree [11] to index the extents of objects and their current velocities. These indexes include the TPR-tree [21] and its variant TPR*-tree [23], which optimize some operations of the TPR-tree. They work by grouping object extents at the reference time into minimum bounding rectangles (MBRs). Figure 2(a) shows the objects $a$, $b$ and $c$ grouped into the same MBR in node $N_1$. Accompanying the MBRs are the velocity bounding rectangles (VBRs), which represent the expansion of the MBRs with time according to the velocity vectors of the constituent objects. The rate of expansion in each direction is equal to the maximum velocity among the constituent objects in the corresponding direction. A negative velocity value implies that the velocity is towards the negative direction of the axis. For example, in Figure 2(a) we can see that the solid arrow on the left of node $N_1$ has a value of -2. This is because the maximum velocity value of the constituent objects in the left direction is 2. Figure 2(b) shows the expanded MBRs at time 1.

The MBR and VBR structure described can be extended by replacing the constituent object extents with smaller MBRs. This when recursively applied creates a hierarchical tree structure. The tree structure is identical to the classic R-tree [11]. The only difference being the algorithms used to insert, delete and query the tree also need to take the velocity information into consideration. The TPR-tree and the TPR*-tree modify the R*-tree's insertion/deletion and query algorithms.

The insertion and deletion algorithms of the TPR*-tree use a cost model proposed by Tao et al. [23] to reduce the expected number of node accesses for a range query $Q$. We briefly describe this cost model below. This cost model is also used by our paper for analyzing the benefits of a partitioned index in Section 4.

Consider a moving tree node $N$ and a moving range query $Q$ for the time interval $[0,1]$ as shown in Figure 3(a). The MBR (VBR) of $N$ is denoted as $N_R = \{N_{R1-}, N_{R1+}, N_{R2-}, N_{R2+}\}$ ($N_V = \{N_{V1-}, N_{V1+}, N_{V2-}, N_{V2+}\}$), where $N_{Ri-}$ ($N_{Vi-}$) is the coordinate (velocity) of the lower boundary of $N$ on the $i^{th}$ dimension, where $i \in \{1, 2\}$. Similarly, $N_{Ri+}$ ($N_{Vi+}$) refers to the upper boundary. MBR (VBR) of $Q$ also can be denoted similar to $N$.

The *sweeping regions* of $N$ and $Q$ are the regions swept by $N$ and $Q$ during the time interval $[0,1]$ (the grey regions shown in Figure 3(a)). To determine whether node $N$ intersects $Q$, we first define the *transformed node* $N'$ with respect to $Q$ as follows: the MBR of $N'$ in the $i^{th}$ dimension is $\langle N_{Ri-} - |Q_{Ri}|/2, N_{Ri+} + |Q_{Ri}|/2 \rangle$; the VBR of $N'$ in the $i^{th}$ dimension is $\langle N_{Vi-} - Q_{Vi+}, N_{Vi+} - Q_{Vi-} \rangle$. To check whether node $N$ intersects $Q$ during the time interval $[0,1]$ is equivalent to checking whether the transformed node $N'$ intersects the center of $Q$ (which is a point) during the time interval $[0,1]$. Therefore, the probability of $N$ intersecting $Q$ (which is the probability of node $N$ being accessed by the query $Q$) during the time interval $[0,1]$ is the same as the probability of $N'$ intersecting the center of $Q$ during the time interval $[0,1]$, which equals to the area of the sweeping region of $N'$ in the time interval $[0,1]$ (the grey region shown in Figure 3(b)). Assuming that the MBR of $Q$ uniformly distributes in the data space and the data space has a unit extent in each dimension. Adding up this probability for every node of the tree, we obtain the expected number of node accesses for the range query $Q$ as:

$$\sum_{every\ node\ N\ in\ the\ tree} V_{N'}(q_T), \quad (1)$$

where $q_T$ is the query time interval; $V_{N'}(q_T)$ is the volume of the sweeping region of $N'$ during $q_T$.

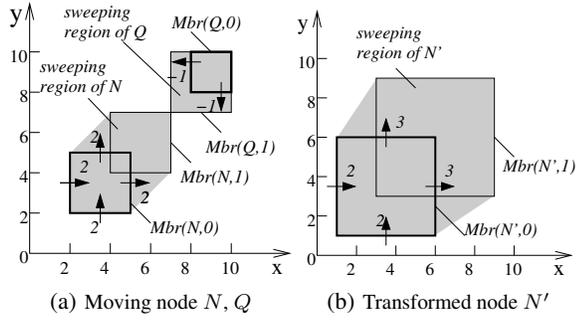

(a) Moving node $N, Q$    (b) Transformed node $N'$

**Figure 3: Sweeping region of moving node**

## 3.2 The $B^x$-tree

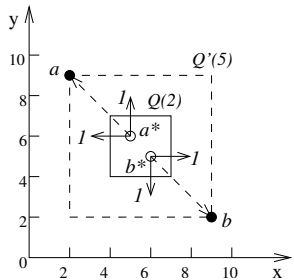

**Figure 4: Query enlargement in the $B^x$-tree**

The $B^x$-tree [13] indexes moving objects using the $B^+$-tree. This is a challenge because the $B^+$-tree indexes 1D space but objects move in a 2D space with associated velocities as well. The $B^x$-tree achieves the challenge by first partitioning the 2D space using a grid, and then using a space-filing curve (Hilbert-curve or Z-curve) to map the location of each grid cell to a 1D space where 2D proximity is approximately preserved. The locations of the moving objects are indexed relative to a common reference time.

The $B^x$-tree incorporates the fact that objects are moving by enlarging the query window according to the maximum velocity of the objects. If the query time is far in the future, and therefore very different from the index reference time, then the query may be enlarged significantly. Figure 4 shows an example of how the window enlargement works. Supposing that the current time is 0, we issue a predictive time slice range query $Q$ at time 2 (the solid rectangle). Considering that moving points $a$ and $b$ (the black dots) stored in the $B^x$-tree, are indexed relative to timestamp 5. From their velocities as shown in Figure 4, we can infer their positions at timestamp 2, which are $a*$ and $b*$ (the circles). The window enlargement technique enlarges the range query $Q$ using the reverse velocities of $a$ and $b$ to get the query window at timestamp 5 (the dashed rectangle). In practice, histograms on a grid base are maintained for the maximum/minimum velocity of different portions of the data space and the query window is enlarged according to the maximum/minimum velocity in the region it covers. Therefore, a drawback of the $B^x$-tree is that, if only a few objects have a high speed, they would make the enlarged query window unnecessarily large for most of the objects.

To reduce the amount of query window enlargement, the $B^x$-tree partitions the index into multiple time buckets, where all objects indexed within the same time bucket are indexed using the same reference time. This results in a smaller difference between the reference time and query time and thus reduces the query window enlargement. When objects are updated, they are moved from the time bucket they are currently residing in to the future time bucket.

## 3.3 Dual Transform Based Moving Object Indexes

The earlier work on dual transform based moving object indexes [1, 16] was improved upon by more recent indexes such as STRIPES [20], the $B^{dual}$-tree [25] and [17]. They index objects in the dual space, i.e. a 4-dimensional space consisting of two dimensions for the location of an object and another two dimensions for the velocity of the object. A consequence of indexing the velocity as separate dimensions is that the moving objects are effectively indexed as stationary objects. All objects are indexed based on the same reference time of 0. A drawback of indexing all objects at the same reference time is that the query search space continues to grow with time, which is overcome by periodically replacing the old index with a new index with an updated reference time.

Dual transform based moving object indexes differ from our work by not exploiting velocity distribution skew to index objects traveling along different *dominant velocity axes (DVAs)* separately.

## 3.4 Indexing Techniques that Handle Skewed Workloads

Zhang et al. [27] propose the $P^+$-tree, which efficiently handles both range and $k$NN queries for different data distributions including skewed distributions. Their work differs from ours in that their index is designed for stationary objects instead of moving objects. Tzoumas et al. [24] propose the QU-Trade technique for indexing moving objects that adapts to varying query versus update distributions by building an adaptive layer on top of the R-tree or TPR-tree. Our work differs from this by adapting to velocity distributions instead of query versus update distributions. Chen et al. [7] propose the ST$^2$B-tree, which improves the $B^x$-tree by making it adaptive to data and query distribution. This is done by dynamically adjusting the reference points and grid sizes. Our work differs from this by creating separate indexes according to velocity distributions instead of adjusting the reference points and grid sizes. Our VP technique can be applied in a straightforward manner to the QUtrade technique and ST$^2$B-tree because their underlying structures are the TPR-tree and the B$^x$-tree, respectively.

Dittrich et al. [8] propose a main memory indexing technique called MOVIES for moving objects. MOVIES assumes that the whole data set resides in memory and the update rate is very high (greater than 5,000,000 per second) whereas our technique does not make such assumptions.

## 3.5 Indexing Techniques for Moving Objects on Networks

There are many existing papers [2, 5, 9, 10] which model the movement of objects along any type of network including road networks. Our paper does not assume that every object must move in a road network, in other words, our technique works for generic scenarios where objects can move freely. Objects moving in road networks is just one of the motivating examples in which case our technique brings great performance gain due to the few dominant directions of object movements.

## 4. HOW VELOCITY PARTITIONING REDUCES SEARCH SPACE EXPANSION

In this section, we analytically show how a velocity partitioned index can reduce the rate of search space expansion. We focus our analysis on the B$^x$-tree and the TPR-tree variants. We first give an intuitive description of a partitioned index versus unpartitioned index. Second, we define search space expansion. Third, we analytically contrast the rate of search space expansion between an unpartitioned index versus a partitioned index. Finally, we present preliminary experimental verification of our analysis.

**Partitioned index.** The main idea of the velocity partitioning (VP) technique is to index objects moving along different DVAs (directions) in separate indexes. It is important to note that the VP technique is not restricted to pairs of DVAs that are perpendicular to each other, but rather will work for any number of DVAs separated by any angle. Here we first use a simple example to illustrate the concept of the VP technique. Later in Section 5, we provide a detailed description of how the VP technique is performed. Figure 5 shows an example of objects indexed by an unpartitioned index versus the same objects indexed by a partitioned index. In this example, objects are moving along two DVAs, the x-axis and the y-axis. In the unpartitioned index, all objects are indexed by the same index. In the partitioned index, objects moving along the x-axis are indexed in a separate index from those moving along the y-axis.

**Search space expansion.** First, we define what we mean by search space expansion. The search space for a query describes the data space that is covered (accessed) when processing the query. The expansion of the search space is determined by the relative movement between the query and the tree nodes. The size of the search space is proportional to the number of tree nodes accessed by a query $Q$, which can be estimated using a cost model proposed by Tao et al. [23] for the TPR-tree/TPR*-tree. The cost model was described in Section 3.1 and given as Equation 1.

Although the cost model was designed for the TPR-tree, it also applies to the B$^x$-tree as follows. For the B$^x$-tree, the query expands but the tree nodes are stationary, which is a special case of the analysis used for Equation 1 where both the query and the tree node are moving and expanding.

The idea behind the cost model of Equation 1 is that we can always transform a moving/expanding query into a stationary one by making relative adjustments to tree nodes. For example, an expanding query and a stationary tree node can be transformed into a stationary query by expanding the tree node by the amount the query was supposed to expand. *Following this line of argument, we only consider the expansion of the tree node in the following analysis without loss of generality.*

Figure 6 shows an example of the search space of the example shown in Figure 5. In the example, $S$ is the search space of the *unpartitioned index*, $S'_X$ and $S'_Y$ are the search space of a *partitioned index* in the $x$- and $y$-axes, respectively. We also assume that all objects are traveling either along the $x$- or $y$-axes, as was the case for Figure 5. The example shows that the search space expands by a quadratic factor for the unpartitioned index versus a linear factor for the partitioned index.

**Analysis of search space expansion of unpartitioned versus partitioned index.** We will first analyze a simplified scenario as shown in Figure 6, and then discuss more general situations in Section 4.1. In this simplified scenario, we assume that: (i) the velocities of all the objects are exactly along the standard $x$- or $y$-axes; (ii) the objects travel in the same speed along all directions; (iii) the extent length of the tree nodes along the $x$- and $y$-axes are the same; and (iv) the initial locations of objects are uniformly distributed in the 2D space. The symbols used in Figure 6 are described as follows. $N'$ is the transformed rectangle of the node $N$ with respect to the query for the unpartitioned index at the initial time 0; $N'_X$ and $N'_Y$ are the transformed rectangles of the node $N$ for the partitioned index for the $x$- and $y$-axes, respectively; $v$ is the maximum speed for the objects in $S$ along both the $x$- and $y$-axes. The extent length of all the nodes is $d$. This assumption is reasonable since we are more interested in the rate of expansion of the search space rather than its initial size.

Let $S'$ denote the combined search space of the partitioned index in the x-axis, $S'_X$ and the y-axis, $S'_Y$ (as shown in Figures 6(b) and 6(c), respectively). Our aim is to show that the rate at which the unpartitioned search space, $S$ expands is higher than the rate at which the partitioned search space $S'$ expands. We quantify the search space as the volume created by integrating the search area from time 0 to the query predictive time $t_h$, where query predictive time refers to the future time of the query. The search area expands with time, therefore we start by expressing the search area of the partitioned index $N'$ as a function of time $t$, $A_{N'}(t)$ as follows:

$$A_{N'}(t) = (d + 2vt)(d + 2vt)$$
$$= d^2 + 4vtd + 4v^2t^2 \quad (2)$$

We are interested in the total expansion of the search area of the partitioned indexed including both the x-axis index and y-axis index. Therefore, let $A_{CN'}(t)$ be the combined area of $N'_X$ and $N'_Y$ as a function of time $t$. $A_{CN'}(t)$ can be computed as follows:

$$A_{CN'}(t) = A_{N'_X}(t) + A_{N'_Y}(t)$$
$$= (d + 2vt)d + d(d + 2vt)$$
$$= 2d^2 + 4dvt \quad (3)$$

We next compute the search volume of $S$. It is important to compute the search volume rather than just the expanded search area since the volume includes the cumulative expansion of the area from time 0 to $t_h$. We compute the search volume $V_S$ of $S$ by integrating the search area $A_{N'}$ from time 0 to $t_h$ as follows:

$$V_S(t_h) = \int_0^{t_h} A_{N'}(t)\, dt$$
$$= \int_0^{t_h} (d^2 + 4vtd + 4v^2t^2)\, dt$$
$$= d^2 t_h + 2dv{t_h}^2 + \frac{4}{3}v^2 {t_h}^3 \quad (4)$$

Similarly the search space volume from time 0 to $t_h$ of $S'$, $V_{S'}$ can be computed as follows:

$$V_{S'}(t_h) = \int_0^{t_h} A_{CN'}(t)\, dt$$
$$= \int_0^{t_h} (2d^2 + 4dvt)\, dt$$
$$= 2d^2 t_h + 2dv{t_h}^2 \quad (5)$$

In order to compare the search space of the partitioned index versus the unpartitioned index, we compute the difference between the search space volume of the partitioned search space $S'$ versus the unpartitioned search space $S$ as a function of time, $\Delta V(t_h)$ as follows:

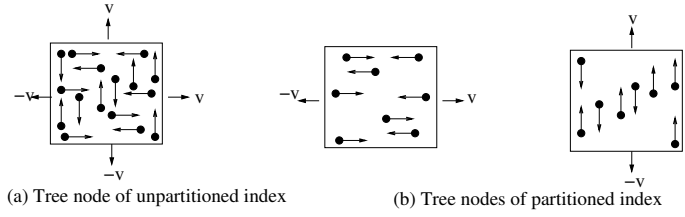

(a) Tree node of unpartitioned index    (b) Tree nodes of partitioned index

**Figure 5: Objects indexed by an unpartitioned index versus the same objects indexed by a partitioned index**

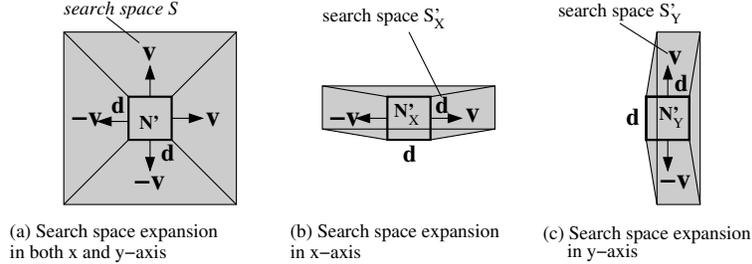

(a) Search space expansion in both x and y−axis    (b) Search space expansion in x−axis    (c) Search space expansion in y−axis

**Figure 6: Search space of unpartitioned index, $S$ versus search space of partitioned index, $S'_X$ plus $S'_Y$**

$$\Delta V(t_h) = V_{S'}(t_h) - V_S(t_h)$$
$$= 2d^2t_h + 2dvt_h{}^2 - (d^2t_h + 2dvt_h{}^2 + \frac{4}{3}v^2t_h{}^3)$$
$$= d^2t_h - \frac{4}{3}v^2t_h{}^3 \quad (6)$$

From Equation 6 we can see that as time increases the search volume of the unpartitioned space $V_S$ becomes increasingly larger than the search volume of the partitioned space, $V_{S'}$. This can be seen by the fact $\Delta V(t_h)$ is negative when $t_h$ is greater than $\frac{d\sqrt{3}}{2v}$. Therefore, when time $t_h$ passes the $\frac{d\sqrt{3}}{2v}$ threshold the search volume of the unpartitioned search volume $V_S$ becomes larger than the partitioned search volume $V_{S'}$.

Next, we analyze the rate of change in the search space, by taking the derivative of Equation 6. This is stated as follows:

$$\frac{d\Delta V(t_h)}{dt_h} = d^2 - 4v^2t_h{}^2 \quad (7)$$

Equation 7 shows that the search volume of the unpartitioned index expands at a much faster rate than the partitioned index. This can be seen by the fact the rate at which the search volume of the unpartitioned index increases above the partitioned index is a squared factor of both $v$ and $t_h$ because $\frac{d\Delta V(t_h)}{dt_h}$ is a squared factor of both $v$ and $t_h$.

The above analysis is with respect to a single node. It obviously applies to any node in the tree and when summing up the search space for all the tree nodes, we reach the conclusion that the query search space on a partitioned index grows much slower with time than the query search space on an unpartitioned index. The following experiment on a real data set validates this result.

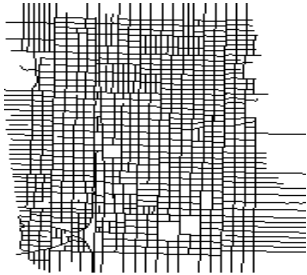

**Figure 8: Chicago road network**

**Experimental verification of the analysis.** Figure 7 shows the results of an experiment, which illustrates the 2D search space expansion for an unpartitioned TPR*-tree and an unpartitioned $B^x$-tree versus a near 1D search space expansion for their partitioned counterparts. The indexes are partitioned using our VP technique (detailed in Section 5). The experiment uses data generated from a portion of the road network of Chicago shown in Figure 8. The experiment involved 100,000 moving objects, with maximum speed of 100 meters per time stamp, with a query predictive time of 60 time stamps. Details of other parameters of the experiment are the default parameters described in the experimental study (Section 6).

Figures 7(a) and 7(b) show the velocity expansion rate of the leaf MBRs for the unpartitioned TPR*-tree and partitioned TPR*-tree, respectively. The results show that the leaf nodes of the unpartitioned TPR*-tree expand in a 2D space whereas the partitioned TPR*-tree expand in a near 1D space. Similarly, Figures 7(c) and 7(d) show the query expansion rate of the unpartitioned $B^x$-tree and partitioned $B^x$-tree, respectively. Again, the query of the unpartitioned $B^x$-tree expands in a 2D space, whereas the partitioned $B^x$-tree expands in a near 1D space.

## 4.1 Discussion of General Cases

In the analysis of the simplified scenario, we have made several assumptions. To lift the first assumption, when the velocities of objects are not exactly along the standard $x$- or $y$-axes, as long as their directions are close to the standard $x$- or $y$-axes, the previous analysis still holds since a small deviation from the *dominant velocity axis (DVA)* incurs a small search space expansion. However, if some objects' directions are not close to any of the DVAs, we will put these objects into an outlier partition. Details of the outlier partition will be discussed in Section 5.2.

An implicit assumption we also made in the previous analysis is that there are two DVAs, one is vertical and the other is horizontal. This assumption may not hold in practice. Therefore, in our *VP* technique, we first find out the actual DVAs (through a combination of PCA and $k$-means clustering). Then, the previous analysis still holds when we replace the $x$- and $y$-axes with the actual DVAs. Details of how to find the DVAs will be discussed in Section 5.1.

## 5. THE VELOCITY PARTITIONING TECHNIQUE

We present our VP technique in this section. Figure 9 shows the system architecture for the VP technique. The system has two main components, a *velocity analyzer* and an *index manager*. The velocity analyzer *partitions a sample* of the velocity of objects from the current workload in order to find the DVAs and an outlier threshold

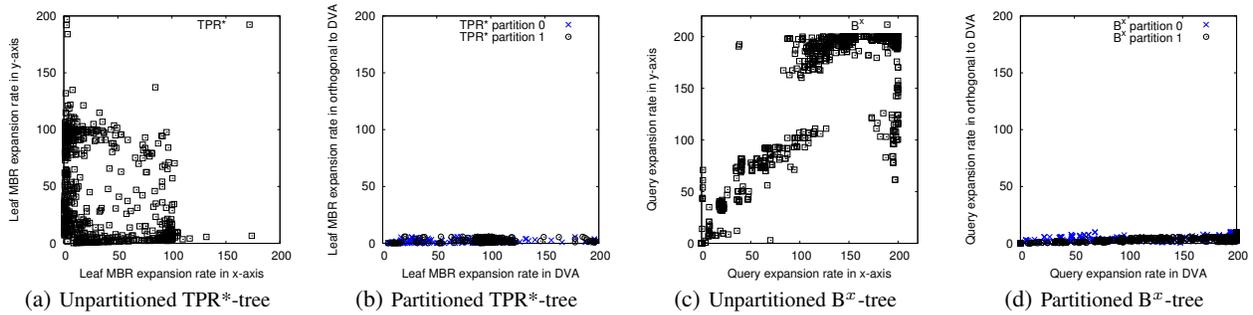

(a) Unpartitioned TPR*-tree  (b) Partitioned TPR*-tree  (c) Unpartitioned $B^x$-tree  (d) Partitioned $B^x$-tree

Figure 7: Search space expansion of the unpartitioned versus partitioned $B^x$-tree and TPR*-tree on the Chicago data set

(used to determine which objects belong to the outlier partition). Velocity is a 2D point in the velocity space, so we refer to the velocity of an object as a *velocity point*. The index manager takes the output of the velocity analyzer to transform the query, insertion and deletion operations to operate on the DVA indexes and outlier index. A DVA index is the same as a traditional moving object index such as the TPR-tree or the $B^x$-tree except objects are indexed using a transformed coordinate space according to the DVA. The index manager inserts an object into the closest DVA index unless it is far from all DVAs, in which case, the object is inserted into the outlier index. If an object update causes its direction of travel to change sufficiently, it may be moved from one index to another. Processing a query involves transforming the query into the coordinate space of each index, and then querying all the indexes and combining the results.

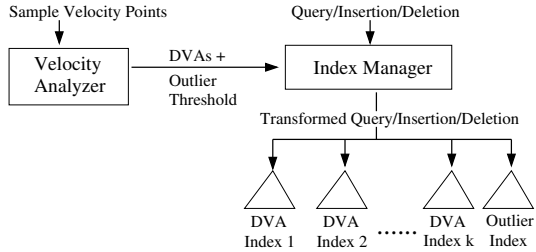

Figure 9: The system architecture of the VP technique

We provide a more detailed description of the velocity analyzer in this section since it is the key component of the system. The velocity analyzer analyzes the sample of velocity points to determine the partition boundaries for future object insertions and querying. The partition boundaries are determined by the DVAs in the data set and an outlier threshold $\tau$. We observe that when there are multiple DVAs in the data set, using only PCA may not be able to identify the DVAs correctly. Therefore, we propose to use a combination of PCA and $k$-means clustering on the sample velocity points to determine the DVAs. Here $k$ is an input value given by the user based on observation of the data set or experience. For example, most road networks have two dominant traffic directions and we can set $k$ to 2. Once the DVAs are determined, the objects can be partitioned based on the closeness of their velocity directions to the directions of the DVAs. However, some velocity points may not be close to any DVA. Those objects are placed in an outlier partition. We determine the boundary of the outlier partition using a threshold $\tau$, which defines an upper bound on what a DVA partition will accept. We choose the $\tau$ value for every partition by analyzing the sample data set using a search space-based cost function.

Algorithm 1 summarizes the *VP* algorithm used by the velocity analyzer. It starts by finding the DVAs using a combination of PCA and $k$-means clustering on the representative sample data (Line 2). Specifically, we integrate PCA into the clustering process itself by using PCA to guide the formation and refinement of clusters. At the end of the clustering process, each cluster contains the velocity points that form one DVA partition. *The 1st PC of each partition is the DVA for the partition.* The partitioning algorithm minimizes the perpendicular distance from each velocity point to the DVAs. The reason we minimize the perpendicular distance is that if all velocity points within one partition have a small perpendicular distance to the DVA, then those velocity points occupy a near 1D space.

We define a *threshold $\tau$* for every DVA to determine whether an object can be accepted to its partition (Line 4). We determine the optimal $\tau$ by minimizing the combined rate of search area expansion of the DVA partition and the outlier partition. Objects whose perpendicular velocity is not within the threshold, $\tau$, of any DVA, are placed in the outlier partition (Line 5). Once all the outlier velocity points have been removed from the DVA partition we recompute the DVA using the remaining velocity points (Line 6). This updated DVA will be a more precise representation of the velocity points now remaining in the DVA partition. The final DVAs and their associated $\tau$ thresholds are used by the index manager for future insertions and query processing.

---

**Algorithm 1:** VelocityPartitioning($A$,$k$)

  **Input**: $A$: sample set of velocity points, $k$: number of DVA partitions
  **Output**: $D$: set of DVAs with associated outlier thresholds $\tau$
1 let $P$ be the set of $k$ DVA partitions with their associated DVAs
2 $P$ = Find DVAs($A$, $k$) // See Algorithm 2
3 **for** *each $p \in P$* **do**
4 | **compute** the maximum perpendicular distance threshold $\tau$ for $p$ according to Section 5.2
5 | **move** the velocity points from $p$ whose perpendicular distance is greater than $\tau$ from the DVA of $p$ into the outlier partition
6 | **recompute** the DVA for the remaining velocity points in $p$
7 let $D$ be the set of DVAs and associated $\tau$ thresholds of $P$
8 **return** $D$

---

In Section 5.1, we describe how our velocity analyzer finds DVAs. In Section 5.2, we describe how our velocity analyzer determines the threshold $\tau$ to decide which objects should be placed in the outlier partition. In Section 5.3, we show how our index manager handles insertion, deletion and update operations. In Section 5.4, we show how our index manager performs the range query. Finally in Section 5.5, we discuss the issue of changing velocity distributions.

### 5.1 Velocity Analyzer: Finding Dominant Velocity Axes (DVAs)

In this subsection, we will first examine two naïve approaches to finding DVAs, and then present our approach for finding DVAs.

**Naïve approach I: PCA.** The first naïve approach is to apply PCA on a sample set of velocity points to find the DVAs. Using PCA to find DVAs is intuitive, since the 1st PC (as described in Section 2.2) represents the principal axis along which the data points lay. In our case, the data points are *velocity* points, therefore, the 1st PC represents the principal axis along which objects travel. However, this approach effectively combines the multiple DVAs in the data set into one average velocity axis, which does not represent any of the individual DVAs. PCA is only useful for finding the DVA when there is only one DVA in the data set. Figure 10(a) shows the result of applying PCA on a sample of 10,000 velocity points

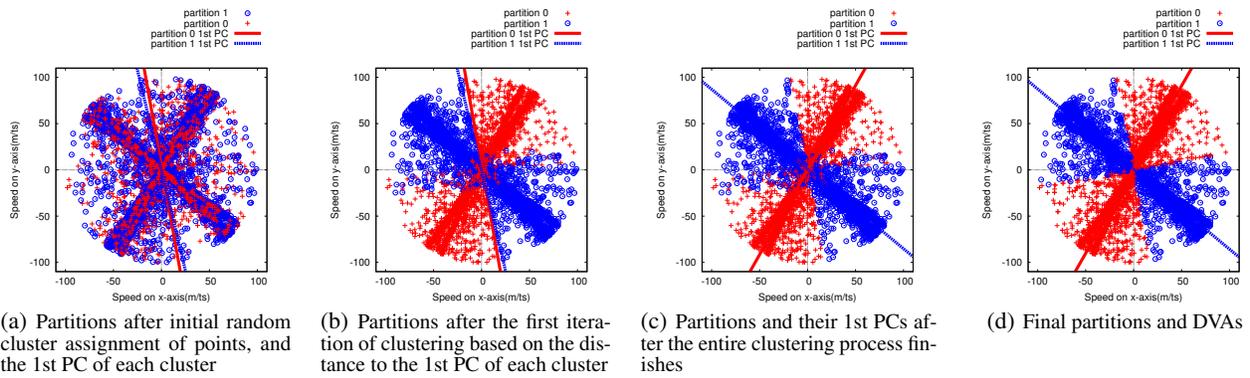

(a) Partitions after initial random cluster assignment of points, and the 1st PC of each cluster

(b) Partitions after the first iteration of clustering based on the distance to the 1st PC of each cluster

(c) Partitions and their 1st PCs after the entire clustering process finishes

(d) Final partitions and DVAs

Figure 11: Our partitioning algorithm being applied to the San Francisco data set shown in Figure 1

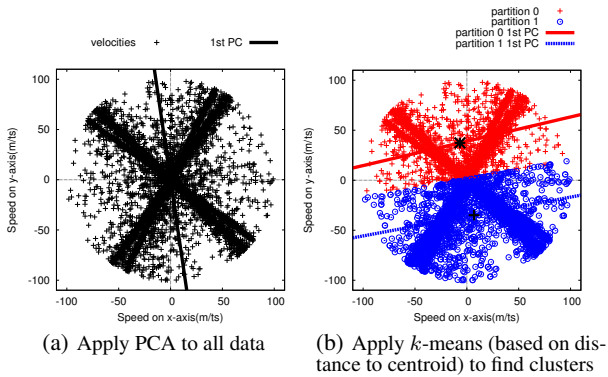

(a) Apply PCA to all data

(b) Apply $k$-means (based on distance to centroid) to find clusters

Figure 10: Result of applying the two naïve approaches to finding the DVAs for the San Francisco data set

of cars traveling on San Francisco network (shown in Figure 1). In this case, the data set has two DVAs but the 1st PC is the average of the two, instead of the two individual DVAs. The 1st PC is far from either of the DVAs. The 2nd PC is orthogonal to the 1st PC and also does not correspond to any of the DVAs.

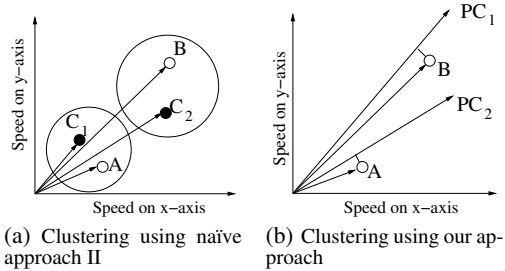

(a) Clustering using naïve approach II

(b) Clustering using our approach

Figure 12: Naïve approach II versus our approach

**Naïve approach II: $k$-means clustering based on distance to centroid followed by PCA on each cluster.** The second naïve approach applies $k$-means clustering to the *velocity* points based on distance to a cluster centroid and then use PCA on each resultant cluster to create one DVA per cluster. This does not work well since it groups objects based on their closeness to a *point* (cluster centroid) rather than closeness to an *axis* (dominant axis). Figure 12(a) shows an example of clustering based on distance to centroid. In the example there are two cluster centroids $C_1$ and $C_2$ and two objects $A$ and $B$. The direction of travel of object $B$ is more aligned to $C_1$ than $C_2$, however the clustering algorithm groups object $B$ with $C_2$ since $B$ is closer to $C_2$. Similar observations can be made for object $A$. Figure 10(b) shows the resultant clusters and corresponding DVAs found on the San Francisco dataset when using $k$-means clustering where distance to centroid is used as the distance measure. Note that the two DVAs found (two parallel lines in Figure 10(b) labeled as 1st PC of partition 0 and 1) by this technique do not resemble the two dominant axes (two axes with the highest concentration of data points) of the data set. The reason is the clusters created center around the cluster centroids shown in Figure 10(b) instead of the dominant axes.

**Our approach: $k$-means clustering based on distance to the 1st PC of each cluster.** In our approach, we use $k$-means clustering on the velocity points, like the naïve approach II, but we use the perpendicular distance to the 1st PC of each cluster (partition) as the distance measure, instead of distance to a centroid. This allows objects to be clustered based on their direction of travel. Figure 12(b) shows an example of using our clustering approach, where there are two clusters with their 1st PCs being $PC_1$ and $PC_2$, respectively. Our algorithm allocates object $A$ to the cluster corresponding to $PC_2$ because $A$ has a shorter perpendicular distance to $PC_2$. Similarly, object $B$ is placed in the cluster corresponding to $PC_1$. This assignment of objects to clusters makes sense since the direction of travel for object $A$ is more aligned to $PC_2$ than $PC_1$, similarly for object $B$.

---

**Algorithm 2:** FindDVAs($A$, $k$)

**Input**: $A$: set of velocity points, $k$: number of partitions
**Output**: $P$: set of partitions with associated 1st PC

1 let $P$ be the set of $k$ partitions
2 **initialize** each partition $p \in P$ to be empty
3 **for** *each velocity point $a \in A$* **do**
4     randomly **assign** $a$ into a partition $p \in P$

5 **while** *at least one velocity point has moved into a different partition* **do**
6     **compute** the 1st PC for each partition in $P$ using PCA
7     **for** *each velocity point $a \in A$* **do**
8         **if** *a is not currently in the partition whose 1st PC has the shortest distance from $a$* **then**
9             **move** $a$ into partition whose 1st PC has the shortest distance from $a$

10 **return** $P$ and associated 1st PC as the DVA partitions and their associated DVAs

---

Algorithm 2 shows precisely how our k-means clustering algorithm based on distance to the 1st PC is used to find DVAs.

Figure 11 shows an example of applying the FindDVAs algorithm with $k = 2$ to the San Francisco data set of Figure 1. Figure 11(a) shows the initial random partitions and their corresponding 1st PCs (Lines 3-4 and 6). Note that although the two initial partitions are randomly created, their two 1st PCs are slightly apart. Next, Figure 11(b) shows the partitions created after reassigning velocity points to their closest 1st PCs. Note that after just this 1st reassignment iteration the partitions already closely resemble the final partitions shown in Figure 11(d). The reason for this is the reassignment of points amplifies the difference between the two 1st PCs by putting points that are slightly closer to one of the 1st PCs in the partition of that 1st PC. Figure 11(c) shows the updated 1st

PC of the partitions after reassigning velocity points (Line 6). The algorithm continues refining velocity points until they converge to the final partitions with their corresponding 1st PC (DVAs) shown in Figure 11(d).

## 5.2 Velocity Analyzer: the Outlier Partition

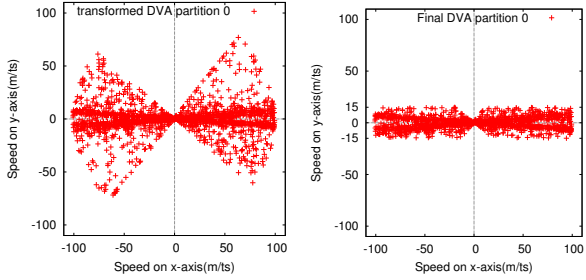

(a) Transformed DVA partition 0    (b) Final DVA partition 0 after removing the outliers

**Figure 13: The transformed DVA partition 0 and its final DVA partition after removing outliers**

Our aim is to have all objects within each partition travelling in a near 1D space. However, from Figure 13(a) we can see that the data points when transformed into the coordinate space formed by DVA 0 of Figure 11 do not travel in a near 1D space, due to the presence of outlier objects. To moderate the influence of these objects, we place those data points with a perpendicular distance above a *threshold* $\tau$ from their DVAs into the outlier partition. A cost analysis is performed upon each DVA partition separately to assign individual $\tau$ values to each DVA partition. The outlier partition is indexed in the standard coordinate system since the objects in it have little correlation with any DVAs.

We determine the optimal $\tau$ value using a slightly simplified version of the search space metric defined at the beginning of Section 4. More specifically we use the minimum total rate of expansion of the area of the transformed leaf nodes $A_{N'_d}$ and $A_{N'_o}$ of the DVA and outlier partitions, respectively. We use the same process as that shown at the beginning of Section 4 to transform the velocities of the queries into the tree nodes. This minimization metric captures the change in the search area as a function of time. We focus our analysis on leaf nodes since non-leaf nodes are typically cached in the RAM buffer, the majority of RAM buffer misses are due to leaf node accesses.

For a given DVA partition and an outlier partition, we define the total rate of expansion of the area of the transformed leaf nodes of the two partitions as follows:

$$\begin{aligned} TA(t, n_d) &= L_d A_{N'_d}(t) + L_o A_{N'_o}(t) \\ &= \frac{n_d}{n_l}(d + 2v_{xmax}t)(d + 2v_{y_d}(n_d)t) \\ &\quad + \frac{(n - n_d)}{n_l}(d + 2v_{xmax}t)(d + 2v_{ymax}t) \end{aligned} \quad (8)$$

where $L_d$ and $L_o$ are the number of leaf nodes in the DVA and outlier partitions, respectively, $n$ is the total number of objects in both partitions, $n_d$ is the number of objects in the DVA partition and $n_l$ is the average number of objects per leaf node. Figure 14 illustrates the other terms used on the equation diagrammatically. The most important term is $v_{y_d}(n_d)$, since this is the term that corresponds to the threshold value $\tau$. $v_{y_d}(n_d)$ is the maximum speed along the $y$-axis in the DVA partition. $v_{y_d}(n_d)$ is a function of $n_d$ as we adjust $v_{y_d}(n_d)$ by removing from the DVA partition the objects whose $y$ component speed is the highest. The remaining terms are described as follows. $d$ is the length along both the $x$- and $y$-axes of both $N'_d$ and $N'_o$. We use the same $d$ for all side lengths because we assume uniform distribution of object locations. $v_{xmax}$ and $v_{ymax}$ are the maximum speed of $N'_o$ along the $x$- and $y$-axes, respectively. For simplicity, we also suppose that the maximum speed of $N'_d$ along the $x$-axis is also $v_{xmax}$. This approximation is reasonable since we partition solely based on the $y$-axis maximum speed and therefore we assume that the maximum speed of object movements along the $x$-axis is approximately the same for all partitions.

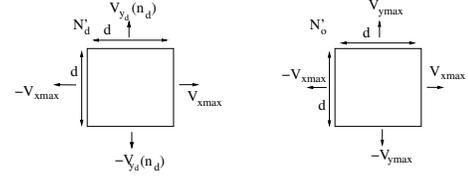

**Figure 14: Diagram used to illustrate the terms used in Equation 8**

Next, we take the derivative of $TA(t, n_d)$ with respect to $t$ to quantify the rate of expansion of $TA(t, n_d)$:

$$\begin{aligned} \frac{d\,TA(t,n_d)}{dt} &= \frac{2n_d}{n_l}((v_{y_d}(n_d) - v_{ymax})(d + 4v_{xmax}t)) \\ &\quad + \frac{2n}{n_l}(dv_{ymax} + v_{xmax}(d + 4v_{ymax}t)) \end{aligned} \quad (9)$$

We need to minimize Equation 9 in order to minimize the rate of $TA(t, n_d)$ expansion. The only components of the equation that are not constant are $n_d$ and $v_{y_d}(n_d)$. Therefore, minimizing Equation 9 is same as minimizing the following expression:

$$n_d(v_{y_d}(n_d) - v_{ymax}) \quad (10)$$

**Algorithm for determining optimal $\tau$ value.** To find the $n_d$ value that minimizes Equation 10 analytically, we would need to have an equation describing $v_{y_d}(n_d)$. However, it is hard to find a general form for the $v_{y_d}(n_d)$ equation because it is data distribution dependent. Therefore, we use an equal width cumulative frequency histogram, per DVA partition, to capture the data distribution of $v_{y_d}(n_d)$. Each bucket of the histogram stores the number of velocity points in the DVA whose maximum $y$ speed is the corresponding $y$ speed of the bucket.

Our algorithm finds the $\tau$ threshold, for each DVA partition, by taking a uniform sample of $v_{y_d}(n_d)$ values and computing the corresponding Equation 10 value. The $v_{y_d}(n_d)$ value giving the minimum value for Equation 10 is used as $\tau$. This approach incurs a small computational cost since Equation 10 is simple and can be computed cheaply. Figure 13(b) shows the final DVA partition 0 after removing outliers from the transformed partition shown in Figure 13(a).

Our experimental study (Section 6.1) shows that the algorithm proposed above is able to find a close to optimal perpendicular distance $\tau$ value for both the $B^x$-tree and the TPR*-tree.

## 5.3 Index Manager: Insertion, Deletion and Update

The insertion algorithm is relatively straightforward. First, the algorithm finds the DVA index $i_{min}$ whose perpendicular distance from the object $o$ is the smallest. Then, if the perpendicular distance of $o$ to $i_{min}$ is larger than $\tau$, then $o$ is inserted into the outlier index otherwise $o$ is inserted into $i_{min}$. Before an object is inserted into $i_{min}$, $o$ is first transformed into the coordinate space of $i_{min}$ using $i_{min}$'s 1st PC. The transformation process involves a simple matrix multiplication between the coordinates of $o$ and the 1st PC of $i_{min}$.

When performing deletion, the algorithm first finds the partition object $o$ resides in via a simple lookup table, and then uses the base index structure's deletion algorithm to delete the object from its partition. When an object changes its velocity, an update is performed on the index.

An update simply consists of a deletion followed by an insertion. The updated object will be inserted into the closest DVA index which may be different from its original DVA index. If an update involves moving an object from one DVA index to another then both indexes need to be locked at the beginning of the update to ensure a concurrent query on the destination index does not miss the inserted object. This may slightly increase the locking overhead.

## 5.4 Index Manager: Range Queries

**Algorithm 3:** RangeQuery($I$, $q$)

**Input**: $I$: set of all indexes including both DVA indexes and the outlier index, $q$: range query
**Output**: $RS$: result set
1 **for** *each index $i \in I$* **do**
2    **if** *$i$ is a DVA index* **then**
3      **transform** the range of $q$ to the coordinate space of index $i$ using the 1st PC of $i$
4      **create** transformed query $q'$ consisting of a rectangular axis-aligned MBR of the transformed range of $q$
5    **else**
6      $q' = q$ // index $i$ is the outlier index
7    **execute** range query $q'$ on index $i$ and store results in $URS$
8    **filter out** the objects in $URS$, which are not contained in $q$ and **add** the remaining objects into $RS$
9 **return** $RS$

In this subsection, we present the range query algorithm, which can be used for both circular and rectangular range queries. Algorithm 3 details the steps the index manager uses to execute the range query. The index manager needs to query each of the indexes separately and merge the results as the query region may encompass objects from different indexes. Before querying each DVA index, we need to first transform the query range into the coordinate space of the DVA index using the 1st PCs of the DVA index (Line 3). The transformation process involves simple matrix multiplication between the coordinates of the query range and that of the 1st PCs. The transformed ranges are bounded by a rectangular minimum bounding region (MBR), which is axis aligned with the coordinate space of the DVA indexes (Line 4). The transformed query is then executed on the indexes using the query algorithm of the underlying index, such as the B$^x$-tree and the TPR*-tree (Line 7). Finally, the objects in the result are filtered to remove any objects, which are in the MBR of the transformed query but not be in the original query region (Line 8). Note that when querying the outlier index, there is no query transformation needed since the outlier index uses the standard coordinate system (Line 6).

Figure 15(a) shows an example of a circular range query $q$ with radius $r$ before transforming into the coordinate space of a DVA index. It also represents the first and the 2nd PCs of the DVA index. Figure 15(b) shows the transformed query $q'$, which is bounded by an axis aligned MBR in the coordinate space of the DVA index formed by the 1st PCs.

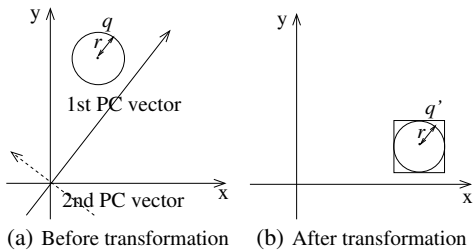

(a) Before transformation     (b) After transformation

**Figure 15: Circular range query before and after transforming into a DVA index's coordinate space**

Our system supports all three query types described in Section 2.1, namely the time slice range query, time interval range query, and moving range query. We discuss the moving range query since it is the most general form of the three query types. After transforming the range query into the transformed coordinate system and applying the filtering step (Line 9 of Algorithm 3), the same object containment relationship with the original query is retained. The query velocity can also be transformed into the new coordinate system and the query can be executed in the standard way. Thus, our system supports the same query types as the underlying indexes (the B$^x$-tree/the TPR*-tree) including the three query types discussed in Section 2.1.

## 5.5 Handling Changing Velocity Distributions

In theory, if the dominant direction of object travel changes significantly we would need to rerun the velocity analyzer to determine new DVAs, and then readjust the indexes to align with the new DVAs. However, we find in real life, the direction component of the velocity distribution changes little since the routes of the moving objects are usually fixed. This is intuitive as velocity distributions are usually dictated by rarely changing environmental factors, such as road networks, flight paths and shipping lanes, etc. Therefore, the dominant direction of object travel is likely to be stable. However, the speed component of the velocity distribution is likely to change with time. For example, during the morning rush hour there will be many cars travelling into the city, resulting in reducing speed. In contrast, during this time, there will be few cars moving out of the city and they will be moving fast. The opposite is true during afternoon rush hour. The speed distribution has no effect on the coordinate system of the DVA indexes since the cars still travel along the same DVA. However, it does affect the value of the threshold $\tau$, since $\tau$ is determined by the $y$-axis speed distribution of objects moving in the transformed coordinate system of the DVA indexes. We handle this situation by continuous updating the histogram used to determine $\tau$, and then periodically computing an updated $\tau$. Computing $\tau$ incurs only a small computational overhead because the equation used to derive it is simple.

## 6. EXPERIMENTAL STUDY

In this section, we report the results of experiments illustrating the performance of our VP technique applied to the B$^x$-tree [13] and the TPR*-tree [23] against their unpartitioned counterparts. We firstly evaluate the ability of our algorithm to find the optimal $\tau$ threshold value. Second, we measure the overhead incurred by the velocity analyzer. Third, we compare both the query and update performance of the algorithms across various data sets. Fourth, we compare the query performance of the algorithms for varying data sizes. Fifth, we measure the effect of varying the maximum speed of object movement. Sixth, we compare the query performance of the algorithms for varying query predictive time. Finally, we show representative results for the rectangular range query.

The experiments were conducted based on the benchmark defined in Chen et al. [6] for evaluating moving object indexes. The road network and synthetic (uniform) data sets used in the experiments were generated using the benchmark's data generator provided by Chen et al. [6]. To generate the road network data sets we fed the road network nodes and edges into the benchmark generator. The road network nodes and edges were all generated using the XML map data from the OpenStreetMap web site (OpenStreetMap.org). We generated four road network data sets. Their characteristics can be summarized as follows:

- The New York (NY) and the Melbourne CBD (MEL) road networks contain the largest number of nodes and edges, and hence average the length of each edge. Therefore, both road networks have the highest update frequency.

- Both the Chicago (CH) and the San Francisco (SA) road networks contain less number of nodes and edges and hence both have smaller number of updates compared to the MEL and the NY networks.

- The CH road network's velocity distribution is the most skewed, followed by the SA, the MEL and the NY road networks.

We focus our experimental study on the circular time slice range query, with a future predictive time ranging from 0 to 120 time stamps as described in Table 1. We focus on the circular query because it resembles many real world occurrences and is also used in the filter step of the $k$ Nearest Neighbor query. The circular range query specifies a range, which is a certain distance from a point. For example, a taxi driver is interested in potential passengers within 200 meters of itself, or a tank wants to know if there are any other tanks within one kilometer of itself. We use the circular range query as the default query. We have performed the same set of experiments for the rectangular range query and the results are similar to those for the circular range query. We show representative results for the rectangular range quer in Section 6.8.

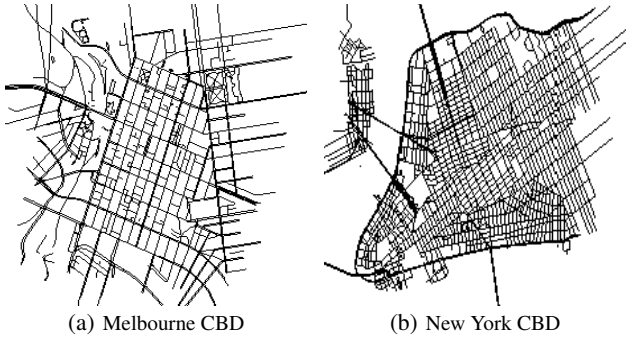

(a) Melbourne CBD     (b) New York CBD

**Figure 16: Other tested road networks**

| Parameter | Setting |
|---|---|
| Space domain (m$^2$) | **100,000x100,000** |
| Cardinality of objects | **100K**, ..., 500K |
| Max. object speed (m/ts) | 20, ..., **100**, ..., 200 |
| Max update interval (ts) | **120** |
| Range query radius (m) | 100,..., **500**,...,1000 |
| Query predictive time (ts) | 0, 10, ..., **60**, ..., 120 |
| Time duration (ts) | **240**, 600 |
| RAM buffer size (pages) | **50** |
| Disk page size | **4KB** |
| Data distribution | **CH**, MEL, SA, NY, uniform |

**Table 1: Parameters and their settings**

The parameters used in the experiments are summarized in table 1, where values in bold denote the default values used.

We compare our VP technique applied on top of two state-of-the-art moving object indexes of contrasting styles: the B$^x$-tree [13] and the TPR*-tree [23] with their unpartitioned counterparts (indexes that has not been velocity partitioned). We used the source code for the TPR*-tree and the B$^x$-tree provided by Chen et al. [6]. All code was implemented in C++ under Microsoft Visual C++ 2008 running on Microsoft Windows 7 Professional SP1. The algorithms compared are described as follows:

- **B$^x$-tree**. The B$^x$-tree [13] has two time buckets and uses the Hilbert curve for space partitioning. We use the improved iterative expanding query algorithm [14] to reduce query enlargement. The histogram used contains 1000x1000 cells.
- **TPR*-tree**. The TPR*-tree [23] is optimized for query size of 1000x1000m$^2$.
- **B$^x$(VP)-tree** and **TPR*(VP)-tree**. The VP technique applied to the B$^x$-tree and the TPR*-tree denoted as B$^x$(VP)-tree and TPR*(VP)-tree, respectively. Both trees use a velocity histogram containing 100 buckets for determining $\tau$ value. We set the number of DVA indexes to 2 because we found that in almost all road network data sets, the roads were aligned to two main axes. The settings for the underlying B$^x$-tree and TPR*-tree are the same as above. The velocity analyzer used for both indexes used 10,000 sample velocity points.

Our experiments measure the following metrics: average I/O per query; average I/O per update; average execution time per query; and average execution time per update. The execution time results include both CPU and I/O time. The update metric results are only reported for one experiment because this paper is focused on improving query performance.

All experiments were conducted on a PC powered by Intel Core i7 CPU 2.8GHz with 8GB DDR3 main memory.

## 6.1 Finding Optimal $\tau$ Threshold

In this experiment, we examine the effectiveness of our algorithm (see Subsection 5.2) at finding the optimal $\tau$ threshold for each index. As mentioned before $\tau$ is used to determine which objects should be placed in the outlier index. We compared the B$^x$(VP)-tree and the TPR*(VP)-tree using different fixed $\tau$ thresholds against the B$^x$(VP)-tree and the TPR*(VP)-tree automatically finding the optimal threshold value according to the algorithm of Section 5.2. We used both the CH and SA road network data sets for this experiment. The results are shown in Figure 17. In Figure 17, the straight lines represent the B$^x$(VP)-tree and the TPR*(VP)-tree using the automatic algorithm for determining $\tau$ and the curves represent the B$^x$(VP)-tree and the TPR*(VP)-tree using different fixed $\tau$ thresholds. The results show that the VP technique is able to automatically compute a near optimal $\tau$ threshold for both real data sets and moving object indexes.

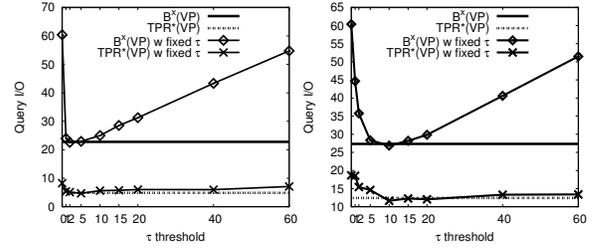

(a) CH road network     (b) SA road network

**Figure 17: $\tau$ algorithm versus varying fixed $\tau$ threshold**

## 6.2 Velocity Analyzer Overhead

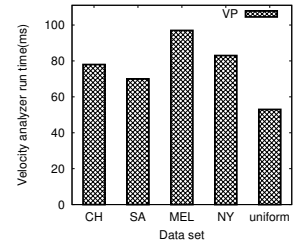

**Figure 18: Overhead of velocity analyzer**

In this experiment, we measure the overhead of running our velocity analyzer as described in Sections 5.1 and 5.2. The velocity analyzer partitions the sample velocity points using a combination of PCA and k-means clustering to arrive at the DVA index boundaries. We performed this experiment across the four road networks, CH, SA, MEL, NY and the uniform synthetic data set. We have run each data set five times and reported the average execution time. The results are shown in Figure 18. The results show that the overhead of the velocity analyzer over all tested data sets is low, taking between 50 milliseconds and 97 milliseconds.

## 6.3 Effect of Varying Data Sets

In this experiment, we compare the algorithms across the four road networks CH, SA, MEL, NY and the uniform synthetic data set. The query I/O and execution time results are shown in Figures 19(a) and 19(b), respectively. The results show that the B$^x$(VP)-tree and the TPR*(VP)-tree consistently outperform their unpartitioned counterparts for road network data sets. The query I/O performance improvement ranges from 280% for the B$^x$-tree on the CH data set to 20% improvement for the TPR*-tree on the NY data set. The performance improvement is due to the fact the VP technique is able to exploit the presence of DVAs in these data sets.

In general, the VP technique is able to improve the query performance of the B$^x$-tree more than the TPR*-tree because the B$^x$-tree does not attempt to group objects travelling in similar directions at all. In contrast, the insertion algorithm of the TPR*-tree attempts to group objects travelling in the same direction into the same tree node, albeit in a locally optimized way instead of the globally optimized way of the VP technique. Therefore, for the TPR*-tree, the performance advantage of using the VP technique is diminished.

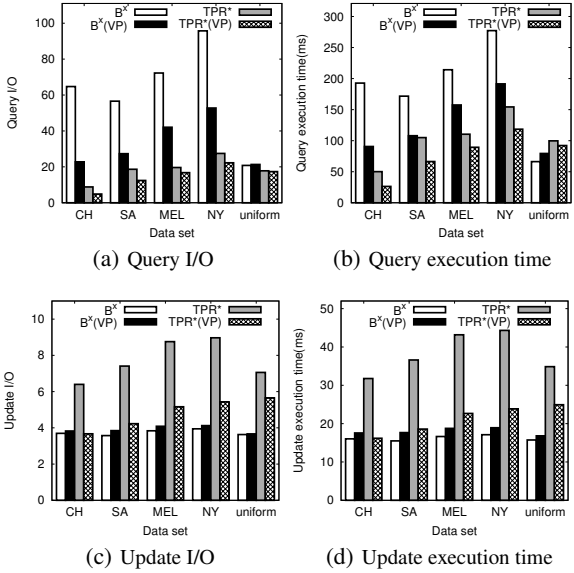

**Figure 19: Effect of varying data sets**

The results for the uniform data set show that the performance advantage of the $B^x(VP)$-tree and the TPR*(VP)-tree over their unpartitioned counterparts is removed. This is because in the uniform data set there are no DVAs, and therefore nothing can be gained from partitioning the index by velocity distributions. In some cases, the $B^x(VP)$-tree performs slightly worse than the unpartitioned counterparts because of the overhead of maintaining multiple indexes and frequently computing an updated $\tau$ threshold.

The update I/O and execution time results for this experiment are shown in Figures 19(c) and 19(d), respectively. The TPR*(VP)-tree outperforms the TPR*-tree by up to a factor of 1.7 for average update I/O cost and up to a factor of 1.9 for average execution time. This is because both the deletion and insertion algorithms of the TPR*-tree involve traversing the tree in a similar fashion to the query. Our algorithm is better at querying than the unpartitioned TPR*-tree. This fact combined with the fact each of the partitioned indexes is smaller than the single unpartitioned TPR*-tree, explains the reason for the faster update performance of the TPR*(VP)-tree compared to the unpartitioned TPR*-tree. However, the update performance of the $B^x(VP)$-tree and the unpartitioned $B^x$-tree are similar. This is because for the $B^x$-tree the update performance is directly proportional to the height of the tree. The height of the $B^x(VP)$-tree and the unpartitioned $B^x$-tree are the same in our experiments. In fact, the $B^x(VP)$-tree is slightly worse than the $B^x$-tree for update performance due to the fact buffering is more effective when there are less trees and the $B^x(VP)$-tree needs to frequently compute an updated $\tau$ threshold.

For the remaining experiments, we only report query cost results and omit the update results because the technique proposed in this paper is mainly aimed at improving the query performance and also we have tight space limitations.

### 6.4 Effect of Data Size on Range Query

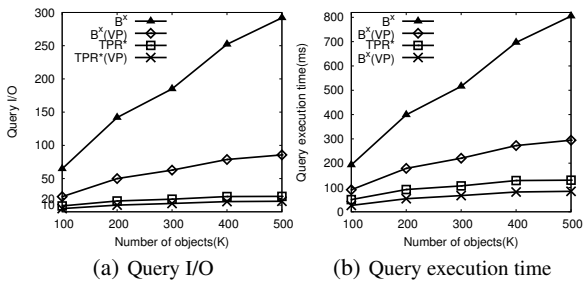

(a) Query I/O  (b) Query execution time

**Figure 20: Effect of data size on range query**

In this experiment, we examine the query performance of each index while varying the number of objects. As the data size grows, Figure 20 shows that the query performance increases approximately linearly across all indexes. We observed that the $B^x$-tree has the worst query performance and scales poorly with increasing number of objects. The results show that the $B^x(VP)$-tree is effective at improving the performance of the unpartitioned $B^x$-tree by up to as much as a factor of 3.4 for I/O and a factor of 2.8 for execution time. The performance improvement of TPR*(VP)-tree over the unpartitioned TPR*-tree is more modest at up to a factor of 1.8 for I/O and 1.9 for execution time. The reason for this is the same as explained in the previous section, namely the TPR*-tree already attempts to group objects moving in the same direction into the same tree node, whereas the $B^x$-tree does not.

### 6.5 Effect of Maximum Object Speed on Range Query

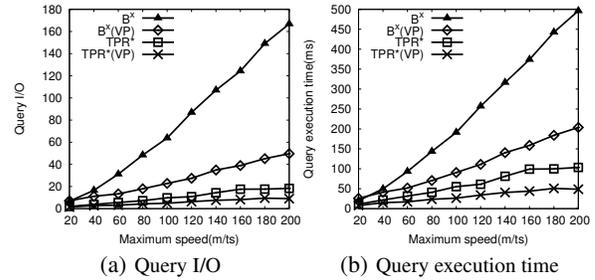

(a) Query I/O  (b) Query execution time

**Figure 21: Effect of maximum object speed on range query**

In this experiment, we study the effect of varying the maximum object speed on the query performance among all the indexes. Figure 21 shows that the $B^x$-tree suffers the most from increases in the maximum object speed and exhibits the steepest increase in both query I/O and query execution time. The reason is that it uses the maximum velocity when expanding queries.

The results show that the VP technique is able to improve the performance of the unpartitioned indexes by an increasing margin as the maximum object speeds increases. This matches the analysis of Section 4.

The $B^x(VP)$-tree outperforms the $B^x$-tree by up to a factor of 3.4 for average query I/O and up to a factor of 2.8 for query execution time. The TPR*(VP)-tree outperforms the TPR*-tree by up to a factor of 2 for average query I/O and up to a factor of 2.1 for query execution time.

### 6.6 Effect of Range Query Size on Range Query

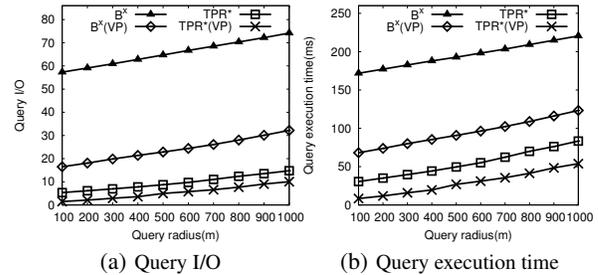

(a) Query I/O  (b) Query execution time

**Figure 22: Effect of range query size on range query**

In this experiment, we vary the radius of the range query. Results in Figure 22 again show that the VP technique is more effective at improving the performance of the $B^x$-tree compared to the TPR*-tree. However, the relative performance difference between the $B^x(VP)$-tree and the TPR*(VP)-tree and their unpartitioned counterparts becomes relatively smaller in percentage terms. The reason for this is that as the query window gets larger the extent size of the query dominates over the query expansion due to the object velocities. The VP technique only reduces query expansion by partitioning the index according to object velocities and does not reduce the query extent size.

More specifically the results show that for a small query size (radius = 100m) the $B^x$(VP)-tree outperforms the $B^x$-tree by up to a factor of 3.5 for query I/O and 2.8 for query execution time and the TPR*(VP)-tree outperforms the TPR*-tree by up to a factor of 3.6 for query I/O and 3.8 for query execution time.

### 6.7 Effect of Query Predictive Time on Range Query

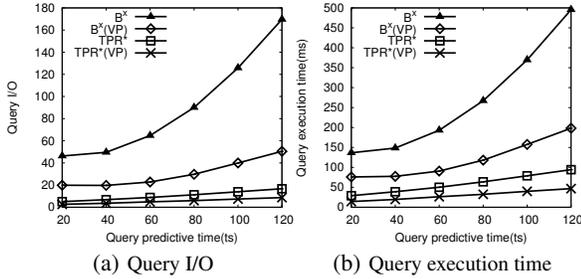

**Figure 23: Effect of query predictive time on range query**

In this experiment, we vary the query predictive time from 20 to 120 time stamps. This experiment is important since it demonstrates how well we can restrict the expansion of the search space as we query further into the future. The results in Figure 23 again show that the query performance of the $B^x$-tree degrades much faster with increasing query predictive time than the other algorithms. Again the VP technique is able to make a large impact on improving the performance of the $B^x$-tree compared to the TPR*-tree. The reasons are similar to the previous experiment, namely the $B^x$-tree expands the query too much but this time due to a larger time value rather than velocity value.

### 6.8 Effect of Query Predictive Time on Rectangular Range Query

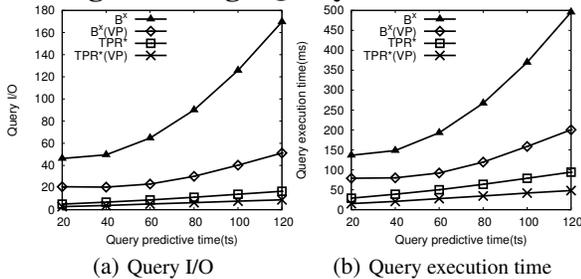

**Figure 24: Effect of query predictive time on the rectangular range query**

As mentioned earlier, we have conducted the same set of experiments for the rectangular range query as the circular range query and the results were similar. However, due to space limitations we only show representative results for the rectangular range query. We choose to vary query predictive time experiment because it tests the ability of the algorithms to handle varying rates of query search space expansion.

In this experiment, the rectangular range queries have side lengths of 1000x1000m$^2$. The results are almost the same as the results for the circular range query.

## 7. CONCLUSION

We have proposed the VP technique, a novel method that improves the moving object index performance by exploiting the skew of velocity distribution. The main idea is to partition objects based on their moving directions, and then use separate indexes to index the objects moving along different dominant velocity axes separately. We first provided analysis to show why this idea should work. Then, we proposed several algorithms to achieve effective velocity partitioning. The VP technique can be applied to most moving object index structures. Finally, we implemented it on two representative index structures, the TPR*-tree and the $B^x$-tree and performed extensive experiments on both real and synthetic data sets. The results show that these index structures equipped with the VP technique outperform their original versions consistently.

## Acknowledgment

This work is supported under the Australian Research Council's Discovery funding scheme (project numbers DP0985451 and DP0880250).